\documentclass[doublespacing]{elsart}

\usepackage{epsfig}
\usepackage{amssymb}

\begin{document}
\begin{frontmatter}
\title{Quantum correlations in spin models}
\author[label1,label2]{Guo-Feng Zhang\corauthref{cor1}},
\corauth[cor1]{Corresponding author:gf1978zhang@buaa.edu.cn}
\author[label2]{Heng Fan},
\author[label2]{Ai-Ling Ji},
\author[label3]{Zhao-Tan Jiang},
\author[label4]{Ahmad Abliz}
\author[label2]{and Wu-Ming Liu}
\address[label1]{Department of Physics, School of Physics and Nuclear Energy Engineering, Beihang University, Xueyuan Road No. 37, Beijing 100191, PR China}
\address[label2]{Beijing National Laboratory for Condensed Matter Physics, Institute of Physics, Chinese Academy of Sciences, Beijing 100190, PR China}
\address[label3]{Department of Physics, Beijing Institute of Technology, Beijing 100081, PR China}
\address[label4]{School of Mathematics, Physics and Informatics, Xinjiang Normal University, Urumchi 830054, PR China}
\begin{abstract}
\quad Bell nonlocality, entanglement and nonclassical correlations are different aspects of quantum correlations for a given state. There are many methods to measure nonclassical correlations.
In this paper, nonclassical correlations in two-qubit spin models are measured by use of measurement-induced disturbance (MID) [Phys. Rev. A, 77, 022301 (2008)] and geometric measure of quantum discord (GQD) [Phys. Rev. Lett. 105, 190502 (2010)]. Their dependencies on external magnetic field, spin-spin coupling, and Dzyaloshinski-Moriya (DM) interaction are presented in detail. We also compare Bell nonlocality, entanglement measured by concurrence, MID and GQD and illustrate their different characteristics.
\end{abstract}

\end{frontmatter}

\section{Introduction}
\quad Quantum correlation arises from noncommutativity of operators representing states, observables, and measurements \cite{sll}. Quantum entanglement, which refers to the separability of the states, is very important in quantum information processing and can be realized in many kinds of physical systems which involve quantum correlation. Quantum correlation seems to have been seldom exploited before. Even many people take it for granted that quantum entanglement is quantum correlation. Now we recognize that quantum entanglement is a special kind of quantum correlation, but not the same with quantum correlation. The most imperfect measurement of entanglement for an arbitrary two-qubit mixed state is the simple procedure derived by Wootters \cite{wk}. Bell nonlocality is usually characterized by the violation of Bell inequalities, which identify the entangled mixed state whose correlations can be reproduced by a classical local mode \cite{ndm}. There is an alternative classifications for correlations, which is based on quantum measurements, has arisen in recent years and also plays an important role in quantum information theory \cite{sll,mpi,nli,slu,nli2}. In particular, quantum discord (QD) \cite{hol} is introduced to measure these quantum correlations. There exists indeed separable
mixed states having nonzero discord, and the separable mixed states can be used to perform useful quantum tasks \cite{ada}. Recently, some authors \cite{twe} have pointed out that thermal quantum discord (TQD), in contrast to entanglement and other thermodynamic quantities, spotlight the critical points associated with quantum phase transitions (QPTs) for some spin chain model even at finite temperature $T$. They think that the remarkable property of TQD is an important tool that can be readily applied to reduce the experimental demands to determine critical points for QPTs. Nevertheless, the evaluation of QD involves optimization procedure, and analytical results are
known only in a few cases. Daki$\acute{c}$ \cite{dakic} \emph{et al}. introduce a geometrical measure of quantum discord (GQD) and derive an explicit expression for the case of two qubits. More recently, Luo \emph{et al}. evaluate the GQD for an arbitrary state and obtain an explicit and tight lower bound, their results show QD actually coincides with a simpler quantity based on von Neumann measurements \cite{ssl3}. They think their simple and explicit bound makes QD a convenient and interesting tool for analyzing quantum correlations. Unlike QD and GQD, Luo \cite{sll} introduced a classical vs quantum dichotomy in order to classify and quantify statistical correlations in bipartite states. They use measurement-induced disturbance (MID) to characterize correlations as classical or quantum. In a word, Bell nonlocality, entanglement and nonclassical correlations are different aspects of quantum correlations, it is desirable to compare these notions of quantum correlations.

\quad In the past, many studies concentrate on entanglement properties of condensed matter systems and application in quantum communication and information. An important emerging field is the quantum entanglement in solid state systems such as spin chains. Spin chains are the natural candidates for the realization of the entanglement compared with the other physical systems. The spin chains not only have useful applications such as the quantum state transfer, but also display rich entanglement features \cite{sbo}. The Heisenberg chain, the simplest spin chain, has been used to construct a quantum computer and quantum dots \cite{dlo}. By suitable coding, the Heisenberg interaction alone can be used for quantum computation \cite{dal,dpd,lfs}. The thermal entanglement, which requires neither measurement nor controlled switching of interactions in the preparing process, becomes an important quantity of systems for the purpose of quantum information. A lot of interesting work about thermal entanglement have been done \cite{man,xwa,glk,kmo,gfz1,yye,dvk,gfz2,xgw,mca}. But quantum correlation in spin chains seems to have been seldom exploited before.

\quad It is very interesting and necessary to study the relation between quantum entanglement and various quantum correlations. Moreover, the effects of external parameters on quantum correlation in spin chain also deserve to be investigated. In this paper, we will explore quantum correlation based on Bell nonlocality, MID and GQD in two two-qubit models. The dependencies of spin-spin coupling, DM interaction and external magnetic field on quantum correlations are investigated. The comparison between quantum correlations and thermal entanglement measured by concurrence will be given.

\quad The paper is organized as follows. In Sec. 2, we recall Bell nonlocality, MID and GQD. In Sec. 3, we will investigate these quantities and quantum entanglement measured by concurrence in two two-qubit spin models and give a detailed comparison. The effects of spin-spin coupling, DM interaction, and external magnetic field are illustrated. Finally, Sec. 4 is devoted to conclusions.

\section{Bell nonlocality, Nonclassical Correlation via MID and GQD, Concurrence}
2.1. Bell nonlocality

\quad It is known that all pure states violate Bell-CHSH inequalities whenever they are entangled. In the case of mixed state, for any two-level systems state $\rho$ we can apply the following criterion \cite{rho1, rho2}, let $m(\rho)=\max_{j<k}\{u_{j}+u_{k}\}$, where $u_{j}(j=1,2,3)$ are the eigenvalues of real symmetric matrix $U_{\rho}$ given by $U_{\rho}=T_{\rho}^{t}T_{\rho}$, the superscript $t$ denotes transpose of matrices, with $T_{\rho}=\{t_{nm}\}$, $t_{nm}=$tr$[\rho\sigma_{n}\otimes\sigma_{m}]$ and $\sigma_{m}(m=1,2,3)$ are the three Pauli matrices. Then $\rho$ violates the Bell-CHSH inequalities if and only if $m(\rho)\geq1$. Here, we use $\max\{0,m(\rho)-1\}$ to quantify the capacity of the state $\rho$ to violate the Bell-CHSH inequality.

2.2. Nonclassical correlation measured by MID

\quad We can apply local measurement $\{\prod_{k}\}$($\prod_{k}\prod_{k^{'}}=\delta_{kk^{'}}\prod_{k}$ and $\sum_{k}\prod_{k}=1$) to any bipartite state $\rho$ (of course, including thermal state). Here $\prod_{k}=\prod_{i}^{a}\otimes\prod_{j}^{b}$ and $\prod_{i}^{a}$, $\prod_{j}^{b}$ are complete projective measurements consisting of one-dimensional orthogonal projections for parties $a$ and $b$. After the measurement, we get the state $\prod(\rho)=\sum_{ij}(\prod_{i}^{a}\otimes\prod_{j}^{b})\rho(\prod_{i}^{a}\otimes\prod_{j}^{b})$ which is a classical state \cite{sll}. If the measurement
$\prod$ is induced by the spectral resolutions of the reduced states $\rho^{a}=\sum_{i}p_{i}^{a}\prod_{i}^{a}$ and $\rho^{b}=\sum_{j}p_{j}^{b}\prod_{j}^{b}$, the measurement leaves the marginal information invariant and is in a certain sense the least disturbing. In fact, $\prod(\rho)$ is a classical state that is closest to the original state $\rho$ since this kind of measurement can leave the reduced states invariant. One can use any reasonable distance between $\rho$ and $\prod(\rho)$ to measure the quantum correlation in $\rho$. In this paper, we will adopt Luo's method \cite{sll}, i.e., quantum mutual information difference between $\rho$ and $\prod(\rho)$,  to measure quantum correlation in $\rho$. The total correlation in a bipartite state $\rho$ can be well quantified by the quantum mutual information $I(\rho)=S(\rho^{a})+S(\rho^{b})-S(\rho)$, and $I(\prod(\rho))$ quantifies the classical correlations in $\rho$ since $\prod(\rho)$ is a classical state. Here $S(\rho)=$-tr$\rho$log$\rho$ denotes the von Neumann entropy, and the logarithm is always understood as base $2$ in this paper. So the quantum correlation can be quantified by the measurement-induced disturbance $\texttt{MID}(\rho)=I(\rho)-I(\prod(\rho))$. The non-orthogonal quantum states can not be distinguishied perfectly, one implication is that a
simple projection measure by orthogonal basis will induce the collapse and decoherence
of the measured non-orthogonal states.
The resultant states are orthogonal and are similar as the classical case. The difference
between initial and the finsal states can be understood as the quantum coherence.
Similarly for a bipartite system measured by corrrelated projection, the classical correlation in the
resultant state can be remained. And MID can be understood as the quantum correlation.

2.3. Nonclassical correlation measured by GQD

\quad The geometric measure of quantum discord(GQD) quantifies the nonclassical correlation by using the nearest Hillber-Schmidt distance between the given state \cite{dakic}. For any two-qubit state
$\rho=\frac{1}{4}[\textbf{1}\otimes\textbf{1}+\sum_{i=1}^{3}(x_{i}\sigma_{i}\otimes\textbf{1}+\textbf{1}\otimes y_{i}\sigma_{i})+\\ \sum_{i,j=1}^{3}t_{ij}\sigma_{i}\otimes\sigma_{j}]$, its geometric measure of quantum discord is evaluated as $
\texttt{GQD}(\rho)=\frac{1}{4}(\|\textbf{x}\|^{2}+\|\emph{\textbf{T}}\|^{2}-k_{max}),
$
where $\textbf{x}=(x_{1},x_{2},x_{3})^{t}$ is a column vector, $\|\textbf{x}\|^{2}=\sum_{i}x_{i}^{2}$, $\emph{T}=\{t_{ij}\}$ is a matrix, and $k_{max}$ is the largest eigenvalue of the
matrix $\textbf{x}\textbf{x}^{t}+\emph{T}\emph{T}^{t}$. Here the superscript $t$ denotes transpose of vectors or matrices.

2.4. Entanglement quantified by concurrence

\quad The entanglement of two qubits state $\rho$ can be measured by the concurrence $C(\rho)$ which is written as $C(\rho)=\max[0,2 \max[\lambda_{i}]-\sum^{4}_{i}\lambda_{i}]$\cite{shi}, where
$\lambda_{i}$ are the square roots of the eigenvalues of the
matrix $R=\rho S\rho^{*}S$, $\rho$ is the density matrix,
$S=\sigma_{1}^{y}\otimes\sigma_{2}^{y}$ and $*$ stands for the
complex conjugate. The concurrence is available no matter whether $\rho$ is pure or mixed.

\section{Bell nonlocality, Nonclassical correlation and Concurrence in two two-qubit spin models}

\quad In this section, we will investigate quantum correlations (Bell nonlocality, GQD, MID and concurrence) in a two-qubit Heigenberg \emph{XXZ} spin model under an inhomogeneous magnetic field and in a two-qubit \emph{XXX} spin model with DM anisotropic antisymmetric interaction. The effects of spin-spin coupling, DM interaction and external magnetic field on these prominent characteristics of quantum physics are illustrated. Also, we will compare these quantities and demonstrate their different properties.

\subsection{Quantum correlation in an XXZ spin model}

\quad First, we consider the model $\emph{H}=\frac{1}{2}[J(\sigma_{1}^{x}\sigma_{2}^{x}+\sigma_{1}^{y}\sigma_{2}^{y})
+J_{z}\sigma_{1}^{z}\sigma_{2}^{z} +(B+b)\sigma_{1}^{z}+(B-b)\sigma_{2}^{z}]$, here $J$ and $J_{z}$ are the real coupling coefficients and $\sigma_{i}^{\alpha}(\alpha=x,y,z; i=1,2)$ are Pauli spin operators. $B\geq0$ is restricted, and the magnetic fields on the two spins have been so parameterized that \emph{b} controls the degree of inhomogeneity. Note that we are working in units so that $B$, $b$, $J$ and $J_{z}$ are dimensionless. The thermal concurrence has been studied in Ref.\cite{gfz1}, here we mainly focus on Bell nonlocality, GQD and MID. The comparison between various quantum correlations will be given. In the following, we will consider the thermal state of this system in equilibrium at temperature $T$, which can be expressed $\rho(T)=e^{-\beta\emph{H}}/Z$, where $\beta=1/(k\emph{T})$ , $k$ is the Boltzmann constant and
$Z=$tr$e^{-\beta\emph{H}}$ is the partition function. For
simplicity, we write $k=1$.

\quad We can obtain thermal state in the standard basis
$\{|1,1\rangle,|1,0\rangle,|0,1\rangle,|0,0\rangle\}$, which can be expressed as
$\rho=\frac{1}{Z}[e^{-\frac{J_{z}+2B}{2T}}|1,1\rangle\langle1,1|+e^{-\frac{J_{z}-2B}{2T}}|0,0\rangle\langle0,0|+\rho_{22}|1,0\rangle\langle1,0|+\rho_{33}|0,1\rangle\langle0,1|-s|0,1\rangle\langle1,0|-s|1,0\rangle\langle0,1|]$,
with
$\rho_{22}=e^{J_{z}/(2T)}(m-n)$, $\rho_{33}=e^{J_{z}/(2T)}(m+n)$, $m=\cosh[\eta/T]$, $n=b\sinh[\eta/T]/\eta$, $\eta=\sqrt{b^{2}+J^{2}}$, $s=e^{J_{z}/(2T)}J\sinh[\eta/T]/\eta$, and $Z=(1+e^{2B/T}+2me^{(J_{z}+B)/T})e^{-(J_{z}+2B)/(2T)}$. We have written the Boltzmann constant $k=1$. Note that we work in units where $J$, $J_{z}$, $B$ and $b$ are dimensionless and $T$ is in unit of the Boltzmann constant $k$.
\begin{figure}
\begin{center}
\epsfig{figure=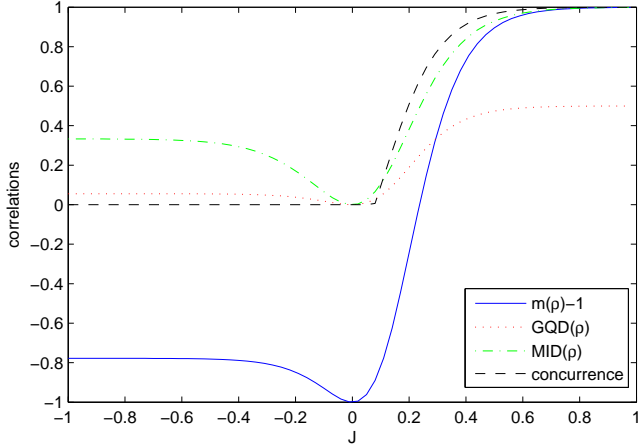,width=0.70\textwidth}
\end{center}
\caption{(Color online) Quantum correlations for \emph{XXX} model without magnetic field when temperature \emph{T}=0.2.}
\end{figure}
\begin{figure}
\begin{center}
\epsfig{figure=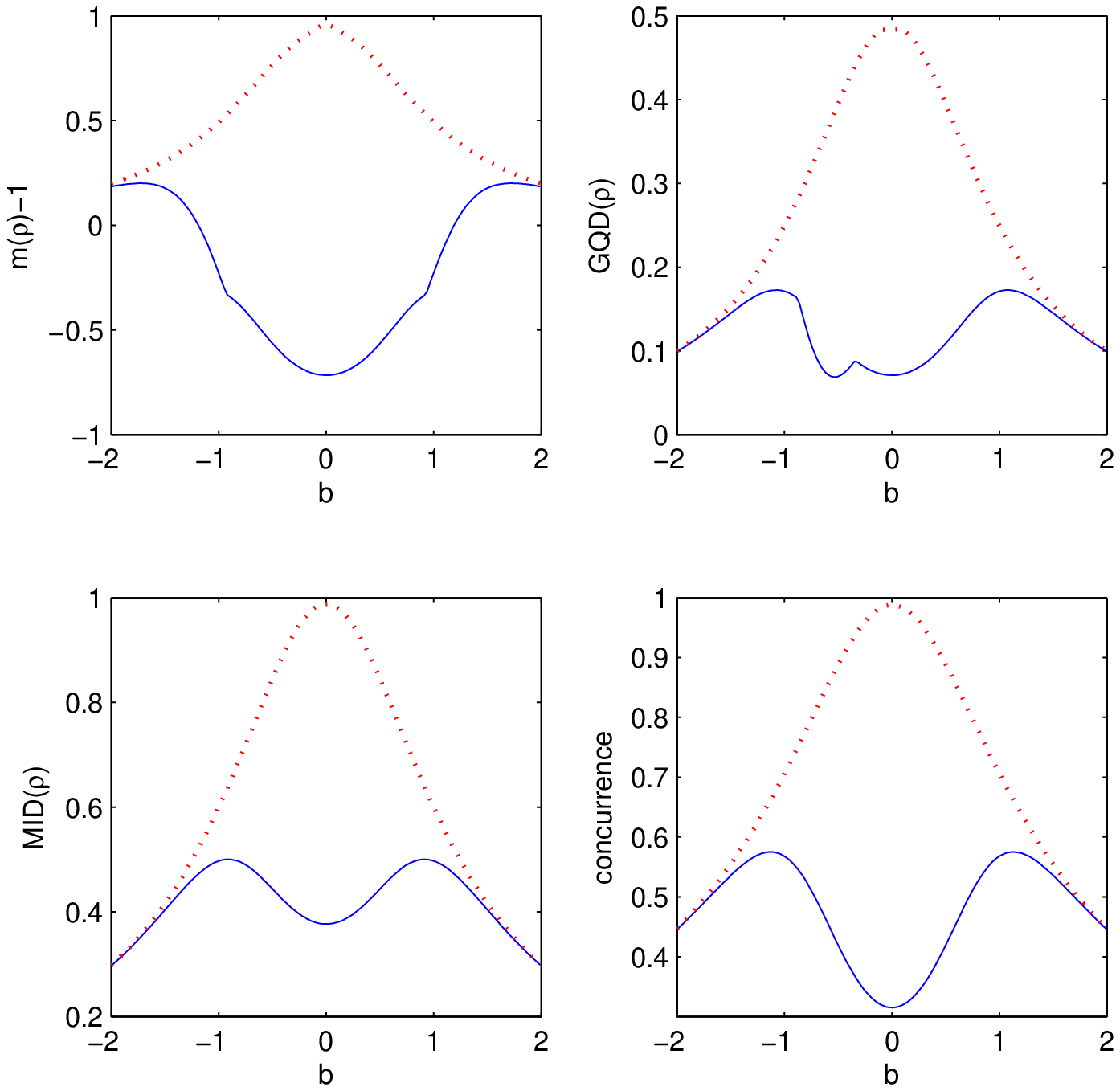,width=0.70\textwidth}
\end{center}
\caption{(Color online) Quantum correlations for \emph{XXZ} model when uniform magnetic field $B=0.6$, temperature $T=0.2$ and spin-spin coupling $J=1$. Solid blue line for $J_{z}=-0.5$ and dotted red line for $J_{z}=0.5$.}
\end{figure}
\begin{figure}
\begin{center}
\epsfig{figure=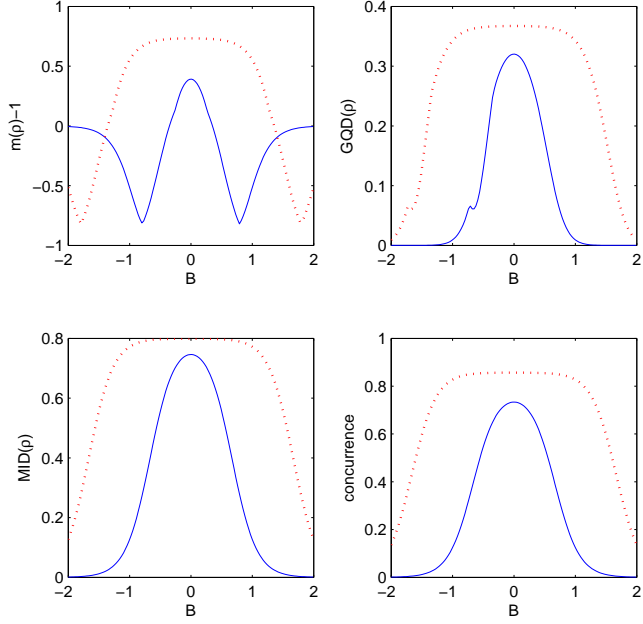,width=0.70\textwidth}
\end{center}
\caption{(Color online) (Color online) Quantum correlations for \emph{XXZ} model when nonuniform magnetic field $b=0.6$, temperature $T=0.2$ and spin-spin coupling $J=1$. Solid blue line for $J_{z}=-0.5$ and dotted red line for $J_{z}=0.5$.}
\end{figure}

\quad \emph{Case 1:}$J_{z}=J$, $B=b=0$. Our model corresponds to an \emph{XXX} spin model. We can easily obtain the relation between GQD and Bell nonlocality as follows.
\begin{equation}
\texttt{GQD}(\rho)=\frac{1}{2(1-2\cot[J/T])^{2}}=\frac{1}{4}m(\rho).
\end{equation}
Because of tedious expression of concurrence and MID, we will not write them here. Figure 1 shows these quantities variation with respect to coupling strength $J$. We can see that antiferromagnetic coupling $(J>0)$ can endure more quantum correlation. Although for $J<0$, there is no entanglement, GQD and MID exist. Moreover, for some entangled states (for example, $J=0.2$, the system is in an entangled state), Bell nonlocality can not be held $(m(\rho)-1<0)$.

\quad \emph{Case 2:} \emph{XXZ} model with magnetic field. We know that in any solid state construction of qubits, there is always the possibility of inhomogeneous Zeeman coupling. So it is necessary to investigate various  quantum correlations's dependencies on magnetic field. In figure 2 and figure 3, the plots of these four quantities with respect to nonuniform magnetic field and uniform magnetic field are given respectively. It is seen that Bell nonlocality, MID and concurrence are symmetric with respect to the zero magnetic field, GQD is not a symmetric function about magnetic field. The introducing of antiferromagnetic coupling $J_{z}>0$ can excite more quantum correlations and entanglement. Quantum correlations evolve alike except for Bell nonlocality.

\subsection{Quantum correlation in an XXX spin model with DM anisotropic antisymmetric interaction}
\quad Next, we consider the model $ H_{DM}=\frac{J}{2}[(\sigma_{1}^{x}\sigma_{2}^{x}+\sigma_{1}^{y}\sigma_{2}^{y}+\sigma_{1}^{z}\sigma_{2}^{z})+\overrightarrow{D}\cdot(\overrightarrow{\sigma_{1}}\times\overrightarrow{\sigma_{2}})]$,
with $J$ is the real coupling coefficient and $\overrightarrow{D}$ is the DM vector coupling. The DM anisotropic antisymmetric interaction arises from spin-orbit coupling \cite{idz,tmo}. For simplicity, we choose $\overrightarrow{D}=D\overrightarrow{z}$. In the standard basis
$\{|1,1\rangle,|1,0\rangle,|0,1\rangle,|0,0\rangle\}$,
thermal state is $\rho=\frac{1}{Z}[e^{-\frac{J}{2T}}|1,1\rangle\langle1,1|+e^{-\frac{J}{2T}}|0,0\rangle\langle0,0|\\+\rho_{22}|1,0\rangle\langle1,0|+\rho_{33}|0,1\rangle\langle0,1|-\frac{M_{-}L_{-}e^{-i\theta}}{2}|0,1\rangle\langle1,0|- \frac{M_{-}L_{-}e^{i\theta}}{2}|1,0\rangle\langle0,1|]$, with $\delta=2J\sqrt{1+D^2}$, $Z=2e^{-J/(2T)}(1+e^{J/T}\cosh[\delta/(2T)])$, $\rho_{22}=\rho_{33}=L_{-}M_{+}/2$, and $L_{\pm}=e^{(J\pm\delta)/(2T)}$, $M_{\pm}=\pm1+e^{\delta/T}$. Using the same method as \emph{3.1}, various quantum correlations can be obtained.

\quad For $D=0$, this model is reduced to an XXX model, the results are the same with the first case in \emph{3.1}. In Ref.\cite{zggann}, the results of MID and concurrence have been given in Fig.6. So, here we only show GQD and Bell nonlocality in Fig.4. We can see that Bell nonlocality and GQD evolve similar but not the same which can be shown clearly from Fig.5. Compared with MID and concurrence in Ref.\cite{zggann}, it is seen that Bell nonlocality and GQD are not more sensitive
to coupling strength $J$ and $D$ than MID and concurrence. Antiferromagnetic coupling can be more helpful for quantum correlations. There is no thermal concurrence for a ferromagnetic \emph{XXX} model when DM interaction is weak. Similarly, Bell nonlocality can not be held when $D$ is small while GQD and MID exist for any DM interaction.
\begin{figure}
\begin{center}
\epsfig{figure=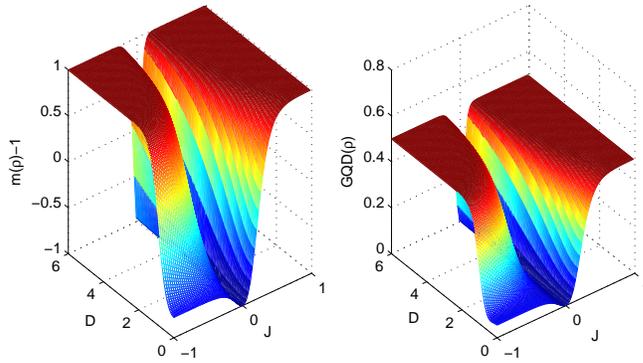,width=0.70\textwidth}
\end{center}
\caption{(Color online) Bell nonlocality and GQD are plotted vs $D$ and $J$ for $T =0.2$.}
\end{figure}
\begin{figure}
\begin{center}
\epsfig{figure=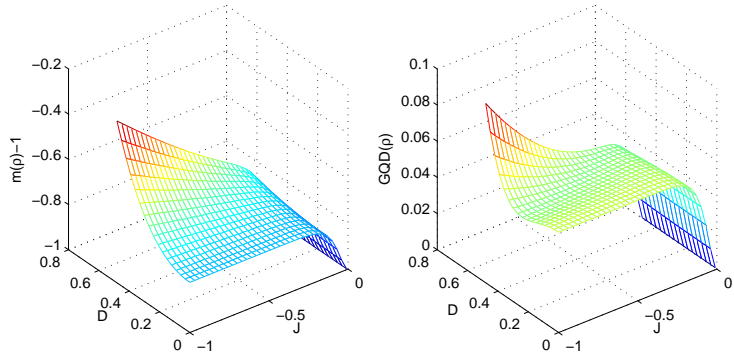,width=0.80\textwidth}
\end{center}
\caption{(Color online) Partial enlarged drawing
of fig.4.}
\end{figure}

\section{Conclusions }
\quad In this paper, we study the evolution of Bell nonlocality, concurrence, and nonclassical correlation measured by GQD and MID respectively. The dependencies of these four quantities on external magnetic field, spin-spin coupling and DM anisotropic antisymmetric interaction are given in detail. More important, we have compared these quantities and found no definite link between them. For an XXX model without magnetic field, antiferromagnetic coupling $(J>0)$ can endure more quantum correlation. Although for $J<0$, there is no entanglement, GQD and MID exist. Moreover, for some entangled states, Bell nonlocality can not be held $(m(\rho)-1<0)$. When magnetic field is present, Bell nonlocality, MID and concurrence are symmetric with respect to the zero magnetic field, GQD is not a symmetric function about magnetic field. The introducing of antiferromagnetic coupling $J_{z}>0$ can excite more quantum correlations and entanglement. Quantum correlations evolve alike except for Bell nonlocality. Moreover, we find that Bell nonlocality and GQD are not more sensitive to coupling strength $J$ and $D$ than MID and concurrence. There is no thermal concurrence for a ferromagnetic \emph{XXX} model when DM interaction is weak. Similarly, Bell nonlocality can not be held when $D$ is small while GQD and MID exist for any DM interaction. We expect our results will be helpful for understanding some related concept of quantum mechanics.

\section{Acknowledgements}
\quad This work was supported by the National Science Foundation of China under Grants No. 10874013, and 10904165. Zhao-Tan Jiang also acknowledges the support of the National Science Foundation of China under Grant No. 10974015. Ahmad Abliz acknowledges the support of the National Science Foundation of China under Grant No. 10664004. Wu-Ming Liu acknowledges the support of NKBRSF of
China under Grants No. 2006CB921400, No. 2009CB930701,
and No. 2010CB922904.

\end{document}